\documentclass[aoas,nameyear,seceqn,dvips]{arximspdf}
\usepackage{graphics}

\doi{10.1214/09-AOAS275}
\volume{3}
\issue{4}
\pubyear{2009}
\firstpage{1776}
\lastpage{1804}

\makeatletter
\def\sfrac#1#2{#1/#2}
\def\vfrac#1#2{(#1)/#2}

\newproclaim{example}{Example}[section]
\makeatother

\begin{document}
\begin{frontmatter}

\title{Semi-parametric dynamic time series modelling with applications
to detecting neural dynamics}
\runtitle{Semi-parametric neural dynamics}

\begin{aug}
\author[a]{\fnms{Fabio} \snm{Rigat}\corref{}\ead[label=e1]{f.rigat@warwick.ac.uk}} and
\author[a]{\fnms{Jim Q.} \snm{Smith}\ead[label=e2]{J.Q.Smith@warwick.ac.uk}}
\runauthor{F. Rigat and J. Q. Smith}
\affiliation{University of Warwick}
\address[a]{Department of Statistics and\\
\quad Centre for Analytical Science\\
University of Warwick\\
Coventry CV4 7AL\\
UK\\
\printead{e1}\\
\phantom{E-mail: }\printead*{e2}} 
\end{aug}

\received{\smonth{10} \syear{2008}}
\revised{\smonth{7} \syear{2009}}

%
\begin{abstract}
This paper illustrates novel methods for
nonstationary time series modeling along with
their applications to selected problems in neuroscience.
These methods are semi-parametric in that
inferences are derived by combining sequential Bayesian
updating with a non-parametric change-point test. As a test statistic,
we propose a
Kullback--Leibler (KL) divergence between posterior distributions
arising from different sets of data. A closed form expression of
this statistic is derived for exponential family models, whereas
standard Markov chain Monte Carlo output is used to approximate its
value and its
critical region for more general models. The behavior of one-step ahead
predictive
distributions under our semi-parametric framework
is described analytically for a dynamic linear time series model.
Conditions under
which our approach reduces to fully parametric state-space modeling are also
illustrated.
We apply our methods to estimating the functional dynamics of a wide
range of neural data, including multi-channel electroencephalogram
recordings, longitudinal behavioral experiments and
in-vivo multiple spike trains recordings. The estimated dynamics are
related to the presentation of visual stimuli, to the evaluation of
a learning performance and to changes in the functional connections
between neurons over a sequence of experiments.
\end{abstract}

%

\end{frontmatter}

\section*{Introduction}

Stochastic modeling of dynamic processes is often implemented
via models having time-dependent parameters
[Hamilton (\citeyear{Hamilton}),
West and Harrison (\citeyear{DLM}),
Fr\"{u}hwirth-Shnatter (\citeyear{SylviaBook})]. For instance, the
coefficients of state-space (SS) and hidden Markov (HM) time series models
[Kalman (\citeyear{Kalman}),
West, Harrison and Migon (\citeyear{WHM85}),
West and Harrison (\citeyear{DLM}),
Cappe, Moulines and Ryden (\citeyear{Cappe})] follow smooth Markovian
processes defined either on their own past or on past values of other latent
variables, whereas those of change-point (CP) models
[Muller (\citeyear{Muller92}),
Stephens (\citeyear{Stephens94}),
Loader (\citeyear{Loader96}),
Mira and Petrone (\citeyear{Anto96}),
B\'elisle et al. (\citeyear{BJMWB}),
Fearnhead and Liu (\citeyear{Fearnhead})] describe pure jump processes.
When these dynamics are specified appropriately,
these time series models can effectively capture nonstationarities
induced by switches
among different dependence regimes
[Hamilton (\citeyear{HamiltonSwitch}),
Shumway and Stoffer (\citeyear{Shum}),
Robert, Celeux and Diebolt (\citeyear{RCD}),
Albert and Chib (\citeyear{Achib}),
McCulloch and Tsay (\citeyear{McCull1}),
Kim (\citeyear{Kim94}),
Ghahramani and Hinton (\citeyear{Gha}),
Fr\"{u}hwirth-Shnatter (\citeyear{Sylvia01})], by smooth changes of the
model parameters through time [Harrison and Stevens (\citeyear{HS76}),
West and Harrison (\citeyear{WH86})] or by the occurrence of
abrupt changes in the data dependence structure [Page (\citeyear{Page55}),
Smith (\citeyear{Smith75}),
Carlin, Gelfand and Smith (\citeyear{CGS}),
Ferger (\citeyear{Ferger95}),
Chib (\citeyear{Chib98})].

This paper illustrates theory and applications of a novel sequential
method for estimating
semi-parametrically the coefficients of time series models having
time-dependent parameters. Our approach is motivated by applications
where little is known about the factors driving the data dynamics.
Here we focus on selected problems in neuroscience where the data
exhibit periods of smooth change interlaced with
occasional large jumps. We model this type of data by combining
sequential Bayesian
updating with a nonparametric change-point test. Sequential
change-point testing is in fact
a well established field which can be traced back at least to the
seminal works
of
Page (\citeyear{Page54}),
Kemp (\citeyear{Kemp57}),
Barnard (\citeyear{Barnard59}) and
Page (\citeyear{Page61}) in statistical process control.
We propose testing for significant changes of a model's parameters
using a novel Kullback--Leibler (KL) divergence
[Kullback and Leibler (\citeyear{KL}),
Kullback (\citeyear{Kbook})]
between their one-step ahead predictive distributions. The null
distribution of this KL
statistic reflects the concentration of the joint posterior
density when new data are generated using the assumed model likelihood
with parameter values drawn from
their current posterior distribution. The semi-parametric nature of our
method stands in
the fact that the value of this test statistic does not depend on the model's
parameters, which are integrated out in the calculation of the KL
divergence.

With respect to the SS and HM families, our approach does not
describe the parameters' dynamics using auxiliary regression
equations depending on known predictors. With respect to CP models, we
do not assume that
parameter values between successive change points are constant.
Instead, we induce a time-dependent parameter process by adopting
different updating strategies depending on whether the
KL statistic lies within its critical region or not. In the former case,
the parameters' joint distribution is updated via Bayes' theorem. In the
latter case, updating is carried out by matching the first two
marginal moments of the current joint posterior probability density to
the prior
for the next time point. This second strategy, which does not
carry the full information content of a posterior distribution
to the future, substantiates the notion
that a change point in the parameter values has been detected.

With respect to SS, HM and CP models, the advantages of our
approach are twofold.
First, in SS and HM models inferences and predictions
are sensitive to the form of the state evolution equations
[Fr\"{u}hwirth-Shnatter (\citeyear{Sylvia}),
Bengtsson and Cavanaugh (\citeyear{Bengtsson})]. Therefore, an exploratory
semi-parametric approach is a natural choice for a first analysis of
time series data when a specific parametrization of the likelihood
function is chosen but no reliable information about the evolution
of its parameters is available
[Robinson (\citeyear{Robinson}),
H\"{a}rdle, L\"{u}tkepohl and Chen (\citeyear{Hardle})]. This is
typically the case for
many biological systems, where dynamic responses to novel
experimental conditions are difficult to anticipate. Second,
the joint distribution of the model's parameters is updated also
between successive
change-points, allowing for a reduction of uncertainty
and for smooth changes of the parameter estimates over time.

From a computational perspective, our approach is motivated by
observing that fully Bayesian sequential inference for a model's
time-dependent parameters and for a
latent multiple change-point process is impractical unless
marginal likelihoods can be calculated explicitly. Otherwise,
the Bayes factors measuring the strength of evidence in the data about
the occurrence of change-points can only be approximated numerically
[Han and Carlin (\citeyear{HanCarlin})]. Current methods for
calculating these approximations require knowledge of
normalizing constants which may be hard to obtain
and they also require estimating
the exact value of a posterior probability density at one point,
which is ideally chosen as one of the posterior modes
[Newton and Raftery (\citeyear{NewtonRaftery}),
Gelfand and Dey (\citeyear{GelfDey}),
Chib (\citeyear{Chib95}),
Chib (\citeyear{Chib98}),
Fr\"{u}hwirth-Shnatter (\citeyear{SylviaBook})].
Our approach represents a practical alternative to these methods in that
point estimates of a latent change-point process are derived without
using marginal likelihoods.

Section~\ref{sec1} of this paper includes its methodological developments.
A general time series framework is introduced and the KL test is illustrated.
A closed form expression of the KL statistic for exponential family
models is derived and examples are presented. Markov chain Monte
Carlo (MCMC) simulation
[Gelfand and Smith (\citeyear{GelfSmith}),
Tierney (\citeyear{Tierney})] is used to approximate
the exact critical region of the KL statistic under
the null hypothesis. This approximation is chosen
as it only requires the assumed
data sampling distribution and the standard MCMC output.
We present a simulation study showing that the power of
the KL change-point test is unaffected in practice by adopting these
MCMC approximations when using a conjugate Bernoulli model.
A sequential algorithm summarizing the computational steps involved in
the implementation of our method is presented.
The behavior of location and spread of the one-step ahead
predictive distributions arising from our method
is described analytically for a conjugate Gaussian linear dynamic model.
Conditions are given so that our semi-parametric approach
reduces to fully parametric state-space dynamic time series modeling.
In Sections~\ref{sec2}, \ref{sec3} and \ref{sec4} our method is applied to estimating three
different types of neural
dynamics. First, we analyze a multivariate time series of
electroencephalogram (EEG) recordings
[Delorme et al. (\citeyear{Delorme}),
Makeig et al. (\citeyear{Makeig})] to reconstruct the time-varying
functional relationships among
different brain areas. Second, we estimate
semi-parametrically a learning curve using a univariate binary time
series arising from a longitudinal behavioral experiment
[Smith et al. (\citeyear{Emery})].
Finally, our method is applied to estimating the functional dynamics of
networks of neurons using in-vivo experimental multiple spike trains
recordings
[Buzs\'aki (\citeyear{Buz})].

\section{Sequential time series modeling and Kullback--Leibler change-point testing}\label{sec1}

Let $\{Y_{i}\}_{i=1}^{N}$ represent a sequence of $K$-dimensional time
series $Y_{i} \in\mathcal{Y}$ of random variables $Y_{i,k,t}$
measured at the time points $t= t_{i,1}<t_{i,2}<\cdots<t_{i,n_{i}}$ with
$t_{i,n_{i}}<t_{i+1,1}$ and $k = 1,\ldots,K$. The distinction between
the $N$ time series is relevant when we allow for the occurrence of time
gaps between them. This situation arises,
for instance, when $N$ consecutive trials are run sequentially
interposed by resting periods. When $n_{i}=1$ for all values of $i$,
we effectively have a single $K$-dimensional time series of length $N$
measured at the time points $t_{i,1}$.
In this paper the time series data $Y_{i}=y_{i}$ are assumed to be
generated by a
finite-dimensional model $P(y_{i} \vert\theta_{i-1},y^{0:(i-1)})$,
such as a vector auto-regressive (VAR) model with shared
coefficients $\theta_{i-1}$ within each of the $N$ periods. The
probability density $f(\theta_{i-1} \vert y^{0:(i-1)})$ here represents
a distribution of the model coefficients
given the initial conditions $y^{0}$ and all past observations up to
and including period $i-1$.
Note that, although we allow the parameter values
to vary in time, neither the functional form of the likelihood
nor the interpretation of its coefficients change over time.

Within this framework, dynamic modeling consists of specifying a
transfer map, taking as arguments the posterior density $f(\theta_{i-1}
\vert y^{0:(i-1)})$, the
time series data $y_{i}$ and possibly other fixed hyper-parameters
$\alpha$ and returning the density $f(\theta_{i} \vert
y^{0:i})$ for $i = 1,\ldots,N$. Various characterizations of
analogous maps are given in
Smith (\citeyear{Jim1,Jim2}). For instance, in standard state-space
models, this transfer map is defined by indexing the prior distribution
for $\theta_{i}$ using the coefficients $\theta_{i-1}$ and a set of
hyper-parameters.
In Markov switching and finite mixture time series models, this
transfer map is again derived by parametric modeling of the joint
density of
the coefficients $\theta_{i}$ and $\theta_{i-1}$ conditional on
the location of a sequence of change-points
[Fr\"{u}hwirth-Shnatter (\citeyear{SylviaBook})].
Here we provide an overview of a transfer map which integrates
sequential Bayesian
inference and change-point testing, leaving to Section~\ref{sec1.1} the
detailed description of an appropriate test statistic.

Let $\hat{\theta}_{i-1}$ be a current point estimate of the model's
parameters at time $t_{i,1}$. When $i=1$ these are prior
summaries, whereas for $i>1$ these estimates incorporate evidence from
past data as described below. If the data $y_{i}$ are generated under
significantly different parameter values with respect to period $i-1$, we
say a change-point has occurred. In this case we define a transfer prior
%
\begin{eqnarray}\label{pmm}
\theta_{i} \sim h(\theta_{i}\vert\hat{\theta}_{i-1}),
\end{eqnarray}
taking as arguments the current parameter estimates and returning a
prior density $h(\cdot)$ for the model's coefficients $\theta_{i}$.
Among the many possible
formulations of this prior, we let its
hyperparameters be functions of the first two marginal moments of the
current posterior
density. Similar forms of prior moment matching have been used for
dynamic point process
modeling by
Gamerman (\citeyear{Gamerman}) and for multi-process dynamic linear
models by
West and Harrison (\citeyear{DLM}).
This partial information transfer from the posterior distribution
ensures that the moment-matched
priors allocate most of their mass around the current marginal
posterior means, but upon detecting a change-point, the dependencies
among different models' parameters, skewness, curtosis and the other
higher-order moments are all reset to their values prior to observing
any data. Equation (\ref{pmm}) represents a partially specified
state evolution density where neither the exact form of the prior
nor the time of occurrence of the change-points are given a
priori. Specific choices for the prior density
depend on the structure of the time series model being entertained and
on the
interpretation of its parameters.

When no change-points are detected
prior to observing the data $Y_{i}=y_{i}$, under (\ref{pmm}) the joint posterior
density of the model's parameters is
%
\begin{equation}\label{posterior}
\hspace*{20pt}f(\theta_{i}\vert y^{0:i},\alpha) \propto
\cases{f\bigl(\theta_{i} \vert y^{0:(i-1)}\bigr)P\bigl(y_{i} \vert\theta_{i},y^{0:(i-1)}\bigr), &\quad\mbox{if }$y_{i} \in\Psi_{i}(\alpha)$,\vspace*{2pt}\cr
h(\theta_{i}\vert\hat{\theta}_{i-1})P\bigl(y_{i} \vert\theta_{i},y^{0:(i-1)}\bigr), &\quad \mbox{if }$y_{i} \notin\Psi_{i}(\alpha)$.
}
\end{equation}
Here $\Psi_{1}(\alpha) = \mathcal{Y}$ and for $i = 2,\ldots,N$ the sets
$\Psi_{i}(\alpha) \subseteq\mathcal{Y}$
include the time series $Y_{i}$ which are inconsistent with their
observed past $y^{0:(i-1)}$ under the current
estimates of the model parameters and the hyper-parameters $\alpha$.

Implementation of (\ref{posterior}) presents two related challenges.
First, it is
essential to formulate the rejection sets $(\Psi_{2}(\alpha),\ldots,\Psi_{N}(\alpha))$ in terms of
a low-dimensional statistic of the data and of the hyper-parameters
$\alpha$. Second, it must be possible to derive the distribution of
such a statistic over the sample space so as to
provide at least a sequential approximation of the rejection sets for
any value of $\alpha$.
A natural way to overcome these challenges is to view
$(\Psi_{2}(\alpha),\ldots,\Psi_{N}(\alpha))$ as the $\alpha$-level
critical regions of a sequential change-point test based on an
appropriate statistic. The transfer map is thus
completely specified by the prior (\ref{pmm})
together with a choice of this test statistic.

\subsection{A Kullback--Leibler change-point statistic}\label{sec1.1}
The Kullback--Leibler divergence
[Kullback and Leibler (\citeyear{KL})] is a well-known information-theoretic
criterion with many applications in statistics, such as density
estimation
[Hall (\citeyear{Hall}),
Hastie (\citeyear{Hastie})], model selection
[Akaike (\citeyear{Akaike2}),
Akaike (\citeyear{Akaike1}),
Carota, Parmigiani and Polson (\citeyear{CPP}),
Goutis and Robert (\citeyear{GRob})], experimental
design [Lindley (\citeyear{Lindley56}),
Stone (\citeyear{Stone})] and the construction of
uninformative priors
[Bernardo (\citeyear{Bernardo})]. Its geometric
properties have been thoroughly explored by
Critchley, Marriott and Salmon (\citeyear{CMS}). The
change-point statistic proposed in this work has a complementary
function to the KL
divergence when used to support model selection.
Instead of testing which
of two competing model structures best predicts one given set of data, here
we construct a statistic detecting whether the same parameter values
could have likely generated two sets of data given a common model
structure.

As change-point test statistic we adopt a Kullback--Leibler divergence
%
\begin{eqnarray}\label{KL}
\operatorname{KL}\bigl(y^{0:(i+1)}\bigr) &=& \displaystyle\int_{\Theta}\log\biggl(\frac{f(\theta_{i}\vert
y^{0:i})}{f(\theta_{i}\vert y^{0:(i+1)})} \biggr)f(\theta_{i}\vert y^{0:i})\,d\theta_{i}\nonumber \\
&=& \log(E_{\theta_{i}\vert y^{0:i}}(P(y_{i+1}\vert\theta_{i},y^{0:i})))
\\
&&{}-E_{\theta_{i}\vert y^{0:i}}(\log(P(y_{i+1}\vert\theta_{i},y^{0:i}))),\nonumber
\end{eqnarray}
where the expectations in (\ref{KL}) are taken with respect to the
posterior density $f(\theta_{i}\vert y^{0:i})$. The
right-hand side of (\ref{KL}) is finite when the
likelihood function is bounded away from zero and infinity for all
values of
the model's parameters and when their posterior density
is proper. In this case (\ref{KL}) is a
nonnegative convex function measuring the discrepancy between the
posterior densities $f(\theta_{i}\vert y^{0:i})$ and
$f(\theta_{i}\vert y^{0:(i+1)})$ over their common support $\Theta$.
Prior to
observing the data $Y_{i+1} = y_{i+1}$, (\ref{KL}) is a random variable
in which
distribution under the null hypothesis depends on that of the future
data $Y_{i+1}$ via the likelihood $P(Y_{i+1}\vert
\theta_{i},y^{0:i})$. The following sections focus on the
interpretation and on the computation of (\ref{KL}).
\subsubsection{Interpretation of the KL statistic and of the change-points}
The scalar hyper-parameter $\alpha$ of the joint
posterior (\ref{posterior}) has the interpretation of the type-1 error
probability for
the change-point test using the statistic (\ref{KL}).
The rejection sets can be written explicitly as intervals $\Psi
_{i}(\alpha) =
(l_{i,\alpha},u_{i,\alpha})$ representing the
$\alpha$-level highest probability interval for the random variable
(\ref{KL})
under the hypothesis of no change over period $i$.

When $\alpha$ is low and (\ref{KL}) lies below the value $l_{i,\alpha
}$, the likelihood
of the observed data is almost a constant in the parameters $\theta_{i}$
over the range of their current posterior density. In other terms, the
parameter values maximizing
the likelihood of the observed $y_{i+1}$ conditionally on the past data
$y^{0:i}$ are given almost zero
probability by the posterior distribution under the hypothesis of
no change. If the change-point statistic lies above
$u_{i,\alpha}$, the parameter values maximizing the likelihood of
the data $y_{i+1}$ are associated to substantial values of the
current joint posterior density, but they are far from its
global maximum. In this case the joint posterior density of
all data $y^{0:(i+1)}$ under the hypothesis of no change is bimodal,
indicating that
the latest batch of data $y_{i+1}$ are not adequately explained by the
current parameter
values. In both cases the value of the
statistic (\ref{KL}) indicates that, in light of the data
$y^{0:(i+1)}$, the
posterior density of the model's parameters
$f(\theta_{i}\vert y^{0:(i+1)})$ significantly departs
from its assumed form under the hypothesis that sequential Bayesian updating
is adequate.

When $\alpha= 0$, no change-point is ever detected, so that the
model's parameters are updated sequentially only via Bayes' rule.
On the other end, if $\alpha= 1$, a~change in the parameter values is
systematically detected
at every time point. In this second limiting case the method proposed
in this work is equivalent to a fully parametric first order Markov
state-space model in which state evolution equations have the form (\ref{pmm}).

\subsubsection{Computation of the change-point statistic}
The test statistic (\ref{KL}) is similar in spirit to the cumulative
Bayes factors
proposed in
West (\citeyear{MikeMonitor}) and
West and Harrison (\citeyear{WH86}), with the
practical advantage that the computation of marginal likelihoods is
not required. However, in general, neither the value of (\ref{KL}) nor
the rejection sets
$(\Psi_{2}(\alpha),\ldots,\Psi_{N}(\alpha))$ may be available in closed
form, so that numerical approximations may be required. In these
cases, at each time period these approximations can be
calculated without incurring in additional computational cost using a
sequence of parameter values $\{\theta^{m}_{i}\}_{m=1}^{M}$ generated
using a Markov chain Monte
Carlo algorithm
[Gelfand and Smith (\citeyear{GelfSmith}),
Smith and Roberts (\citeyear{SR}),
Tierney (\citeyear{Tierney})] having as its target the current
posterior probability
density. Using this technique, the value of (\ref{KL})
is approximated by the average
%
\begin{eqnarray}\label{MCp}
\operatorname{KL}\bigl(y^{0:(i+1)}\bigr) \approx\log\biggl(\frac{\sum_{m=1}^{M}p_{i+1}^{m}}{M} \biggr)-
\frac{\sum_{m=1}^{M}\log(p_{i+1}^{m})}{M},
\end{eqnarray}
where $p_{i+1}^{m} = P(y_{i+1} \vert\theta^{m}_{i},y^{0:i})$ is the
likelihood of the data $y_{i+1}$ given the parameter values
$\theta^{m}_{i}$ and the past data $y^{0:i}$. Using (\ref{MCp}), the
null distribution of (\ref{KL}) can be approximated as follows:
\begin{enumerate}[(ii)]
\item[(i)] for each draw $\theta^{m}_{i}$ generate a pseudo-realization
$y^{m}_{i+1}$ using the joint sampling distribution $P(Y_{i+1} \vert
\theta^{m}_{i}, y^{0:i})$;
\item[(ii)] compute the statistic $\operatorname{KL}(y^{0:(i+1)}_{m})$, where
$y^{0:(i+1)}_{m} =
(y_{0},\ldots,y_{i},y_{i+1}^{m})$, using its Monte Carlo approximation
(\ref{MCp}).
\end{enumerate}
The empirical distribution of the sequence
$\{\operatorname{KL}(y^{0:(i+1)}_{m})\}_{m=1}^{M}$ approximates that of the
$\operatorname{KL}$ statistic (\ref{KL}) under the hypothesis of no change. Therefore,
the empirical
$(\frac{\alpha}{2},1-\frac{\alpha}{2})$th percentiles of the sequence
$\{\operatorname{KL}(y^{0:(i+1)}_{m})\}_{m=1}^{M}$ approximate the rejection sets
$\Psi_{i}(\alpha) = (l_{i,\alpha},u_{i,\alpha})$ for any given
value of $\alpha$.

\subsection{Change-point test power and sample size}\label{sec1.2}

When the time series $\{Y_{i}\}_{i=1}^{N}$ have substantially
different lengths, the power of the change-point test based on the $\operatorname{KL}$
statistic is theoretically unchanged.
For any value of $\alpha$, this invariance is ensured by the
behavior of the posterior distribution at
the denominator of (\ref{KL}).
When the data $Y_{i+1}=y_{i+1}$ carries a large amount of information about
the coefficients of model $P(Y_{i+1} \vert
\theta_{i},y^{0:i})$, their joint posterior distribution under
the hypothesis of no change concentrates by a corresponding large
amount, so that
the distribution of the KL divergence concentrates over large
values. If $Y_{i+1}$ is not expected to carry much additional
information about the model parameters, for instance, due to its small
sample size $n_{i+1}$, the null distribution of the
KL discrepancy is concentrated over small nonnegative values.
This mechanism represents an automatic adaptation of the critical
region $\Psi_{i+1}(\alpha)$ of the KL test, ensuring that its power
does not vary with the
sample size of the data sequentially accrued over time.

%
\begin{figure}[b]

\includegraphics{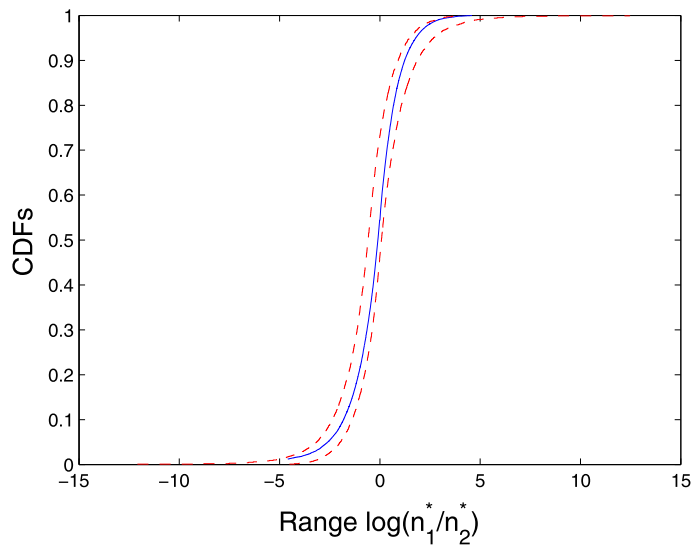}

\caption{The solid line represents the empirical cumulative
distribution function (CDF) of the random variable $Z$ for the 79,743
simulations where
the hypothesis of no change was accepted. The dashed lines represent
the approximate end-points of the point-wise $99\%$ probability intervals
for the CDF of a standard double exponential random
variable. Acceptance of the hypothesis of no change did not cause a significant
departure of the distribution of $Z$ from that of a standard double
exponential distribution,
suggesting that the power of the change-point test is not significantly
affected by different
sample sizes $(n_{1},n_{2})$.}\label{fig1}
\end{figure}

Although this property is sufficiently clear in theory, it is
an open question whether the power of the test is significantly
affected when our method is implemented using the MCMC approximations
outlined above. Here we briefly investigate this issue by simulation using
a conjugate Bernoulli model.
One hundred thousand simulations were run. For each simulation, two
sample sizes $n_{1}$ and $n_{2}$ were
independently generated as independent draws from a discrete uniform
distribution on
the integers $(1,\ldots,M)$ with $M=100$. A success probability $\pi$
was also
independently generated for each simulation using a
uniform distribution on the interval $(0,1)$. Conditionally on
$(n_{1},n_{2},\pi)$,
two independent samples of Bernoulli random
variables were generated, $Y_{1} \sim
\operatorname{Ber}(\pi,n_{1})$ and $Y_{2} \sim \operatorname{Ber}(\pi,n_{2})$. For each simulation,
a sample of size 5000 was generated from the conjugate posterior
$\operatorname{Beta}(1+\sum_{j=1}^{n_{1}}Y_{1,j},1+n_{1}-\sum_{j=1}^{n_{1}}Y_{1,j})$
to compute the Monte Carlo\vspace*{1pt} approximation of the KL statistic and of the
end-points of its $95\%$ probability interval under the hypothesis of
no change. For this simulation study the type-1 error probability of
the test was fixed to $\alpha=0.2$. Under this sampling scheme, with
$n^{*}_{i} = n_{i}-1$ for $i=1,2$, the random variables $\frac
{n^{*}_{1}}{M-1}$ and
$\frac{n^{*}_{2}}{M-1}$ are independent and approximately uniform on $(0,1)$,
so that the distribution of the statistic $Z = \log(\frac
{n^{*}_{1}}{n^{*}_{2}} )$ is
approximately standard double-exponential. If the power of the KL
change-point test is in practice not affected by
the values of $(n_{1},n_{2},\pi)$, the distribution of
$Z$ for the group of simulations where the hypothesis of no change is
accepted should
be standard double exponential. Figure~\ref{fig1} represents with a solid line the
empirical cumulative distribution function (CDF) of $Z$ for the 79,743
simulations where
a significant change was not detected. The two dashed lines in the same figure
represent the point-wise $99\%$ probability intervals
for the CDF of a standard double exponential random variable. Since
at each point the former CDF always lies within its $99\%$ interval,
this simulation study suggests that for the
Bernoulli model the power of the change-point test is not significantly
affected by different
sample sizes $(n_{1},n_{2})$.

\subsection{Sequential fitting and change-point testing algorithm}

This section provides a summary of the computational
steps involved by the dynamic modeling
method illustrated so far. Despite not addressing any
model-specific issues such as the explicit
form of posterior distributions, we aim at providing here a general blueprint
for implementing our method starting from the first sample $y_{1}$:
\begin{enumerate}[(iii)]
\item[(i)] Upon observing the data $y_{1}$, derive the posterior density
\[
f(\theta_{1} \vert y^{0:1}) \propto h(\theta_{1}\vert\hat{\theta
}_{0})P(y_{1}\vert\theta_{1},y_{0}),
\]
where $\hat{\theta}_{0}$ represents an estimate of
the parameter values as reflected by the initial conditions $y_{0}$.
\item[(ii)] Having observed data $y_{2}$, compute the statistic
$\operatorname{KL}(y^{0:2})$ and its rejection interval $\Psi_{1}(\alpha) =
(l_{1,\alpha},u_{1,\alpha})$ as described in Section~\ref{sec1.1}.
\item[(iii)] If $l_{1,\alpha} < \operatorname{KL}(y^{0:2}) < u_{1,\alpha}$, no
change-point is detected. In this case the prior density for
$\theta_{2}$ is the posterior at point (i) and the
posterior density for $\theta_{2}$ derived using Bayes' rule is
\[
f(\theta_{2} \vert y^{0:2},\alpha) \propto f(\theta_{2} \vert
y^{0:1})P(y_{2}\vert\theta_{2},y^{0:1}).
\]
\item[(iv)] Otherwise, match the first two posterior moments
$\hat{\theta}_{1}$ to those of the prior for $\theta_{2}$ and again
apply Bayes' rule, deriving the conditional posterior density
\[
f(\theta_{2} \vert\hat{\theta}_{1}, y^{0:2},\alpha)\propto h(\theta
_{2}\vert\hat{\theta}_{1})P(y_{2}\vert\theta_{2},y^{0:1}).
\]
\end{enumerate}
In case (iv) above, the sequentially estimated change-point process
up to and including
times $(1,2)$ reports one change at time 2. Consistently with the
interpretation of the KL
statistic, the model parameters are updated using all
data starting from the last detected change-point, if any. When a
change is detected at level $1-\alpha$, the new parameter
values are updated using their conditional posterior distribution
under the transfer prior (\ref{pmm}) and the likelihood of the latest
batch of data.

\subsection{Change-point KL statistic for exponential family models}

Several properties of the KL divergence for exponential
family models have been explored by
McCulloch (\citeyear{McCull}).
Here we show that in this circumstance also the divergence~(\ref{KL}) has a closed form. In this case the
algorithm illustrated in Section~\ref{sec1.2} is simplified, as only
the critical intervals $\Psi_{i}(\alpha) =
(l_{i,\alpha},u_{i,\alpha})$ need being approximated. Without loss of
generality, in what follows we assume
that no change-point is detected prior to period $i$. Also, we let
$Y_{i}$ be a 1
dimensional sample of conditionally independent observations with
length $n_{i}$ and joint density [Diaconis and Ylvisaker (\citeyear
{Diaconis79})]
%
\begin{eqnarray}\label{Dia79}
P(Y_{i} \vert\theta_{i}) = \prod
_{j=1}^{n_{i}}a(Y_{i,j})e^{Y_{i,j}\theta_{i} - b(\theta_{i})},
\end{eqnarray}
where $\theta_{i}$ is a scalar canonical parameter.
Diaconis and Ylvisaker (\citeyear{Diaconis79}) show that
each element of $Y_{i}$ has mean and variance
\begin{eqnarray*}
E(Y_{i,j}\vert\theta_{i}) = \frac{\partial b(\theta_{i})}{\partial
\theta_{i}},\qquad
V(Y_{i,j} \vert\theta_{i}) = \frac{\partial^{2}b(\theta_{i})}{\partial\theta_{i}^{'}\,\partial\theta_{i}}.
\end{eqnarray*}
Using the prior
\[
f(\theta_{i} \vert n_{0},y_{0}) = c(n_{0},S_{0})e^{S_{0}\theta
_{i}-n_{0}b(\theta_{i})},
\]
where $S_{0} = n_{0}y_{0}$ for scalar $n_{0}$ and $y_{0}$, the
posterior for $\theta_{i}$ given
the past data $y^{0:i}$ has conjugate density
%
\begin{eqnarray}\label{efpost}
f(\theta_{i} \vert n(i),y^{0:i}) = c(n(i),S(i))e^{n(i) ((S(i)/n(i))\theta_{i}-b(\theta_{i}) ), }
\end{eqnarray}
where\vspace*{1pt} $n(i) = \sum_{j=0}^{i}n_{j}$, $S(i)= \sum_{j=0}^{i}n_{j}\bar
{y}_{j}$ and $\bar{y}_{j}$
represents the arithmetic mean of sample $y_{j}$.
Using the results of
Guti\'errez-Pe\~{n}a (\citeyear{GutPena}), the posterior mean and
variance of $\theta_{i}$ are
\begin{eqnarray*}
E(\theta_{i}\vert n(i),S(i)) = \frac{\partial H(n(i),S(i))}{\partial S(i)},\qquad
V(\theta_{i}\vert n(i),S(i)) = \frac{\partial H(n(i),S(i))}{\partial S(i)^{2}},
\end{eqnarray*}
and the posterior mean and variance of the function $b(\theta_{i})$ are
\begin{eqnarray*}
E(b(\theta_{i})\vert n(i),S(i)) &=& \frac{\partial H(n(i),S(i))}{\partial n(i)},\\
V(b(\theta_{i})\vert n(i),S(i)) &=& \frac{\partial H(n(i),S(i))}{\partial n(i)^{2}},
\end{eqnarray*}
where $H(n(i),S(i)) = -\log(c(n(i),S(i)) )$.
Using these results, we derive the following explicit form for the KL
divergence (\ref{KL}):
\begin{theorem*}
When the posterior density for the coefficients
$\theta_{i}$ has
form (\ref{efpost}), given the data up to and including $y_{i+1}$, the
Kullback--Leibler statistic (\ref{KL}) is
%
\begin{eqnarray}\label{KLef1}
\operatorname{KL}\bigl(y^{0:(i+1)}\bigr) &=& \log \biggl(\frac{c(n(i),S(i))}{c(n(i+1),S(i+1))}\biggr)
-S_{i+1}\frac{\partial H(n(i),S(i))}{\partial S(i)}\nonumber\\[-8pt]\\[-8pt]
&&{}+ n_{i+1}\frac{\partial H(n(i),S(i))}{\partial n(i)},\nonumber
\end{eqnarray}
where the terms on the right-hand side of (\ref{KLef1}) are defined above.
\end{theorem*}

\begin{pf}
By letting the posterior densities $f(\theta_{i}\vert
n(i),S(i))$ and $f(\theta_{i}\vert n(i+1),S(i+1))$ have form (\ref
{efpost}), the KL (\ref{KL})
becomes
\begin{eqnarray*}
\operatorname{KL}\bigl(y^{0:(i+1)}\bigr) =
\log\biggl(\frac{c(n(i),S(i))}{c(n(i+1),S(i+1))} \biggr)-S_{i+1}E(\theta
_{i})+n_{i+1}E(b(\theta_{i})).
\end{eqnarray*}

For exponential family models, the expectations $E(\theta_{i})$ and
$E(b(\theta_{i}))$ with respect to $f(\theta_{i}\vert
y^{0:i})$ are reported above. By
substituting these expressions, equation~(\ref{KLef1}) obtains the
following.
\end{pf}
\begin{example}\label{ex1.1}
When $Y_{i}$ is a Gaussian random
variable with mean $\mu_{i}$ and precision $\lambda_{i}$,
its distribution can be written in the form (\ref{Dia79})
using the two-dimensional statistic
\begin{eqnarray*}
Y^{*}_{i} = [Y_{i},Y^{2}_{i}]
\end{eqnarray*}
and the canonical parameter
\begin{eqnarray*}
\theta_{i} = [\theta_{1,i},\theta_{2,i}] = \biggl[ \lambda_{i}\mu_{i},-\frac
{\lambda_{i}}{2} \biggr]
\end{eqnarray*}
with
\begin{eqnarray*}
a(Y^{*}_{i}) &=& (2\pi)^{-\sfrac{1}{2}},\\
b(\theta_{i}) &=& -\frac{1}{2}\log(\theta_{2,i})-\frac{\theta
_{1,i}^{2}}{\theta_{2,i}}.
\end{eqnarray*}
The conjugate prior for $(\mu_{i},\lambda_{i})$ is Normal-Gamma
$N(\mu_{i} \vert\gamma,\lambda_{i}(2\alpha-1))\operatorname{Ga}(\lambda_{i}\vert
\alpha,\break\beta)$ with coefficients $\alpha> 0.5,\beta>0,\gamma\in
\mathcal{R}$ and normalizing constant
[Bernardo and Smith (\citeyear{BBernardo})]
\[
c(n_{0},S_{0}) = \biggl(\frac{2\pi}{n_{0}} \biggr)^{\sfrac{1}{2}}\frac{S_{2,0}^{\sfrac{S_{1,0}}{2}}/2}{\Gamma(\vfrac{n_{0}+1}{2} )},
\]
where $n_{0} = 2\alpha-1$, $y^{*}_{0} = [y^{*}_{1,0},y^{*}_{2,0}] =
[\gamma,\frac{2\beta}{2\alpha-1}+\gamma^{2}]$, $S_{1,0} =
n_{0}y^{*}_{1,0}$ and $S_{2,0} = n_{0}y^{*}_{2,0}$. Upon observing
the realization $(y_{1},\ldots,y_{i})$, the normalizing constant of the
corresponding conjugate posterior is
\[
c(n(i),S(i)) = \biggl(\frac{2\pi}{n(i)} \biggr)^{\sfrac{1}{2}}\frac{S(2,i)^{\sfrac{S(1,i)}{2}}/2}{\Gamma(\vfrac{n(i)+1}{2} )},
\]
where $n(i) = n_{0}+i$, $S(1,i) = S_{1,0} + \sum_{j=1}^{i}y_{j}$ and
$S(2,i) = S_{2,0} + \sum_{j=1}^{i}y_{j}^{2}$. When also
$y_{i+1}$ is observed, using (\ref{KLef1}), the KL statistic can be
written as
\begin{eqnarray*}
\operatorname{KL}\bigl(y^{0:(i+1)}\bigr) &=&
\log\biggl(\Gamma\biggl(\frac{n(i+1)+1}{2} \biggr)\Big/\Gamma\biggl(\frac{n(i)+1}{2} \biggr)\biggr)
+\frac{1}{2}\log\biggl(\frac{n(i+1)}{n(i)} \biggr)
\\
&&{}+ \log\biggl(\frac{S(2,i)^{\sfrac{S(1,i)}{2}}}{S(2,i+1)^{\sfrac{S(1,i+1)}{2}}} \biggr)
-\frac{y_{i+1}}{2}\log\biggl(\frac{S(2,i)}{2} \biggr)\\
&&{}
-y_{i+1}^{2}\frac{S(1,i)}{S(2,i)}+\frac{1}{2n(i)}+\Gamma\biggl(\frac{n(i)+1}{2} \biggr)
\frac{\partial\Gamma(\vfrac{n(i)+1}{2} )}{\partial n(i)}.
\end{eqnarray*}
\end{example}
\begin{example}
Let $Y_{i}$ be a sample of size $n_{i}$ of
conditionally independent Bernoulli random variables with
success probabilities $\{\pi_{i}\}_{i=1}^{N}$. The canonical
representation of
the Bernoulli probability mass function obtains, by letting $\theta_{i} =
\log(\frac{\pi_{i}}{1-\pi_{i}} )$, $b(\theta_{i}) =
\log(1+e^{\theta_{i}} )$ and $a(Y_{i}) = 1$. The conjugate
prior for $\pi_{i}$ is $\operatorname{Beta}(S_{0},m_{0})$, where
$m_{0} = n_{0}-S_{0}$. Upon observing $(y_{1},\ldots,y_{i})$, the conjugate
posterior
is $\operatorname{Beta}(S(i),m(i))$, where $S(i) = \sum_{j=0}^{i}S_{j}$,
$n(i) = \sum_{j=0}^{i}n_{j}$\vspace*{1pt} and $m(i) = n(i)-S(i)$.
When also $y_{i+1}$ is observed, the KL statistic (\ref{KLef1}) has form
\begin{eqnarray*}
\operatorname{KL}\bigl(y^{0:(i+1)}\bigr) &=& \log\biggl(\frac{\prod_{k=1}^{n_{i}}(n(i)+k)\prod
_{w=1}^{n_{i}-S_{i}}(m(i)+w)}{\prod_{j=1}^{S_{i}}(S(i)+j)} \biggr)
\\
&&{}- S_{i+1}\frac{\Gamma(S(i))}{\Gamma(m(i))}
\frac{\partial(\Gamma(m(i))/\Gamma(S(i)))}{\partial S(i)}
\\
&&{}
+n_{i+1}\biggl(\frac{\partial(\Gamma(n(i))\Gamma(m(i)))}{\partial n(i)}\biggr)\Big/(\Gamma(n(i))\Gamma(m(i))),
\end{eqnarray*}
where $m(i) = n(i)-S(i)$.
\end{example}

\begin{example}
Let $Y_{i}$ represent the random number of
events of a given kind observed within a time interval
$(t_{i,1},t_{i,n_{i}}]$ of fixed length. For this example we assume that
the latter is identical for all samples $i=1,\ldots,N$.
Let the random times at which the events take
place be distributed according to a homogeneous Poisson process with
intensity $\lambda_{i}$, so that the distribution of $Y_{i}$ is Poisson
with parameter $\lambda^{*}_{i}=\lambda_{i}(t_{i,n_{i}}-t_{i,1})$.
The canonical form of the Poisson distribution has parameter
$\theta_{i} = \log(\lambda^{*}_{i})$ and functions $a(Y_{i}) =
\frac{1}{Y_{i}!}, b(\theta_{i}) = e^{\theta_{i}}$. The conjugate prior
for $\lambda^{*}_{i}$
is Gamma with parameters $\operatorname{Ga}(S_{0},n_{0})$ having mean $y_{0}$ and
variance $\frac{y_{0}}{n_{0}}$.
Upon observing $(y_{1},\ldots,y_{i})$, the conjugate posterior for
$\lambda^{*}_{i}$ is $\operatorname{Ga}(S(i),n(i))$ with $S(i) = S_{0}+\sum
_{j=1}^{i}y_{j}$, $n(i) =
n_{0}+i$. When also $y_{i+1}$ is observed, using (\ref{KLef1}), the KL
statistic has form
\begin{eqnarray*}
\operatorname{KL}\bigl(y^{0:(i+1)}\bigr)
&=& \log\biggl(\frac{S(i)n(i)^{S(i)}}{n(i+1)^{S(i+1)}} \biggr)
\\
&&{}+y_{i+1} \biggl(\log(n(i))-\biggl(\frac{\partial\Gamma(S(i))}{\partial S(i)}\biggr)\big/\Gamma(S(i)) \biggr)-\frac{S(i)}{n(i)}.
\end{eqnarray*}
\end{example}
%
\subsection{Effect of change-points on predictive densities}

In this section we illustrate analytically the effect of detecting a
change-point on the one-step ahead predictive density using the
transfer prior (\ref{pmm}) and a conjugate Gaussian dynamic linear model.
For each value of $i$, in what follows we let the scalar random
variable $Y_{i}$ be distributed as $N(\mu_{i},\sigma_{i}^{2})$.
Analogously to Example~\ref{ex1.1}, the prior distribution
for $\theta_{i} = (\mu_{i},\sigma_{i}^{2})$ is taken as the conjugate
Normal-inverse Gamma
\begin{eqnarray*}
\mu_{i} &\sim& N(\hat{\mu}_{i^{*}-1},\sigma_{i}^{2}),\\
\sigma_{i}^{2} &\sim& \operatorname{IGa}\biggl(\frac{\nu}{2},\frac{\nu}{2}\hat{\sigma}_{i^{*}-1}^{2}\biggr),
\end{eqnarray*}
where $1 \leq i^{*}<i$ is the time of the last detected
change-point and $(\hat{\mu}_{i^{*}-1}, \hat{\sigma}_{i^{*}-1}^{2})$ represent
the estimated mean and variance of the joint posterior density at time $i^{*}$.
If $i^{*}=1$, $(\mu_{0},\sigma_{0}^{2})$ represents a fixed initial
condition.
Here the prior density of the variance is
%
\[
f(\sigma_{i}^{2} \vert\nu,\hat{\sigma}_{i^{*}-1}^{2}) =
\frac{(\nu\hat{\sigma}_{i^{*}-1}^{2}/2)^{\sfrac{\nu}{2}}}
{\Gamma(\sfrac{\nu}{2})}\sigma_{i}^{-2(\sfrac{\nu}{2}+1)}e^{-(\nu\hat{\sigma}_{i^{*}-1}^{2})/(2\sigma_{i}^{2})}.
\]
Under this formulation, the prior expectation of the mean is $\hat{\mu
}_{i^{*}-1}$ and that of the variance
is $\frac{\sfrac{\nu}{2}}{\sfrac{\nu}{2}-1}\hat{\sigma}_{i^{*}-1}^{2}$.
If follows that the one-step ahead marginal
predictive distribution is a noncentral Student-$t$. In absence of
change-points prior to time $i$, the predictive density is
%
\begin{eqnarray}\label{pd1}
Y_{i+1} \sim t_{\nu+ i} \biggl(\tilde{\mu}_{i},\frac{i+1}{i+2}\frac{\nu
+i}{2}\frac{1}{\tilde{\sigma}^{2}_{i}} \biggr),
\end{eqnarray}
where
\begin{eqnarray*}
\tilde{\mu}_{i} &=& \frac{1}{1+i}\mu_{0}+\frac{i}{1+i}\bar{y}^{(1:i)},\\
\tilde{\sigma}^{2}_{i} &=& \frac{\nu}{2}\sigma^{2}_{0}+\frac
{i}{2}s^{2}_{(1:i)}+\frac{i}{i+1}\bigl(\mu_{0}-\bar{y}^{(1:i)}\bigr)^{2},
\end{eqnarray*}
and $(\bar{y}^{(1:i)},s^{2}_{(1:i)})$ represent respectively the
sample mean and variance of the data $y^{1:i}$.
If a change-point is detected by the KL statistic (\ref{KL}) at time
$1<i^{*}<i$, under the transfer prior (\ref{pmm}), the conditional
predictive density is
%
\begin{eqnarray}\label{pd2}
Y_{i+1}
\sim t_{\nu+ i - (i^{*}-1)} \biggl(\tilde{\mu}^{*}_{i},\frac
{i-i^{*}+2}{i-i^{*}+3}\frac{\nu+i-i^{*}+1}{2}\frac{1}{(\tilde{\sigma
}^{2}_{i})^{*}} \biggr),
\end{eqnarray}
where
\begin{eqnarray*}
\tilde{\mu}^{*}_{i} &=& \frac{1}{i-i^{*}+2}\hat{\mu}_{i^{*}-1}+\frac
{i-(i^{*}-1)}{i-i^{*}+2}\bar{y}^{(i^{*}:i)},\\
(\tilde{\sigma}_{i}^{2})^{*} &=& \frac{\nu}{2}(\hat{\sigma
}_{i^{*}-1})^{2}+\frac{i-(i^{*}-1)}{2}s_{(i^{*}:i)}^{2}+\frac
{i-(i^{*}-1)}{i-i^{*}+2}\bigl(\hat{\mu}_{i^{*}-1}-\bar{y}^{(i^{*}:i)}\bigr)^{2}.
\end{eqnarray*}
Since the mean and variance of the noncentral Student-$t$
random variable with density $t_{\nu}(\mu,\sigma^{2})$ are
respectively equal to $\mu$ and to $\frac{\nu}{\nu+2}\sigma^{2}$,
equations (\ref{pd1}) and (\ref{pd2}) provide a characterization of
the one-step ahead posterior predictive moments as a function of the time
of the last detected change-point and of the inverse-Gamma prior
coefficient $\nu$. For $i^{*}>1$ the predictive mean is less influenced
by the
sample mean of the data preceding the change-point,
$\bar{y}^{1:(i^{*}-1)}$, and it is more heavily influenced by
$\bar{y}^{i^{*}:i}$, that is, the sample mean
of the data from the change-point on. When a
change-point is detected, the predictive variance is larger with
respect to the case of no change. Its relative increase is a
decreasing function of the difference $(i-i^{*})$, which measures
how far in time the change-point occurred, and it is an
increasing function of the coefficient $\nu$, which measures the
strength of the prior at the initial time.

This behavior is consistent with the intuition that predictions
ensuing from a dynamic time series model should discount
the information content of remote data and focus on more recent
data when significant dynamics occur. In absence of an autoregressive
model structure, as in the
present section, the distinction between remote and recent data is
entirely left to the timing of the detected change-points.

\section{Analysis of multivariate EEG recordings}\label{sec2}

This section presents an application of the methods discussed
above to estimating neural functional dynamics using
multivariate electroencephalogram (EEG) recordings. The data analyzed
here arise from a sequence of 80 identical tests
each having length of approximately four seconds with sampling rate
of 128 points per second. During each test, the same subject
was to press a button when a green square appeared in a specific
screen location
[Makeig et al. (\citeyear{Makeig})]. Previous analyses of these data have
emphasized aspects of time-dependent interactions among different
EEG channels, such as an increased overall synchronization of different brain
areas after presentation of the visual stimulus
[Delorme et al. (\citeyear{Delorme})].
The multidimensional EEG time series are modeled here as a
discrete time Gaussian stochastic process, in which randomness is
thought of as
arising from the intrinsic variability of the brain
activity and from the presence of experimental artifacts.
We describe the dynamic functional relationships among different brain areas
using the time-dependent means and
covariance matrices indexing the data likelihood.

\begin{figure}

\includegraphics{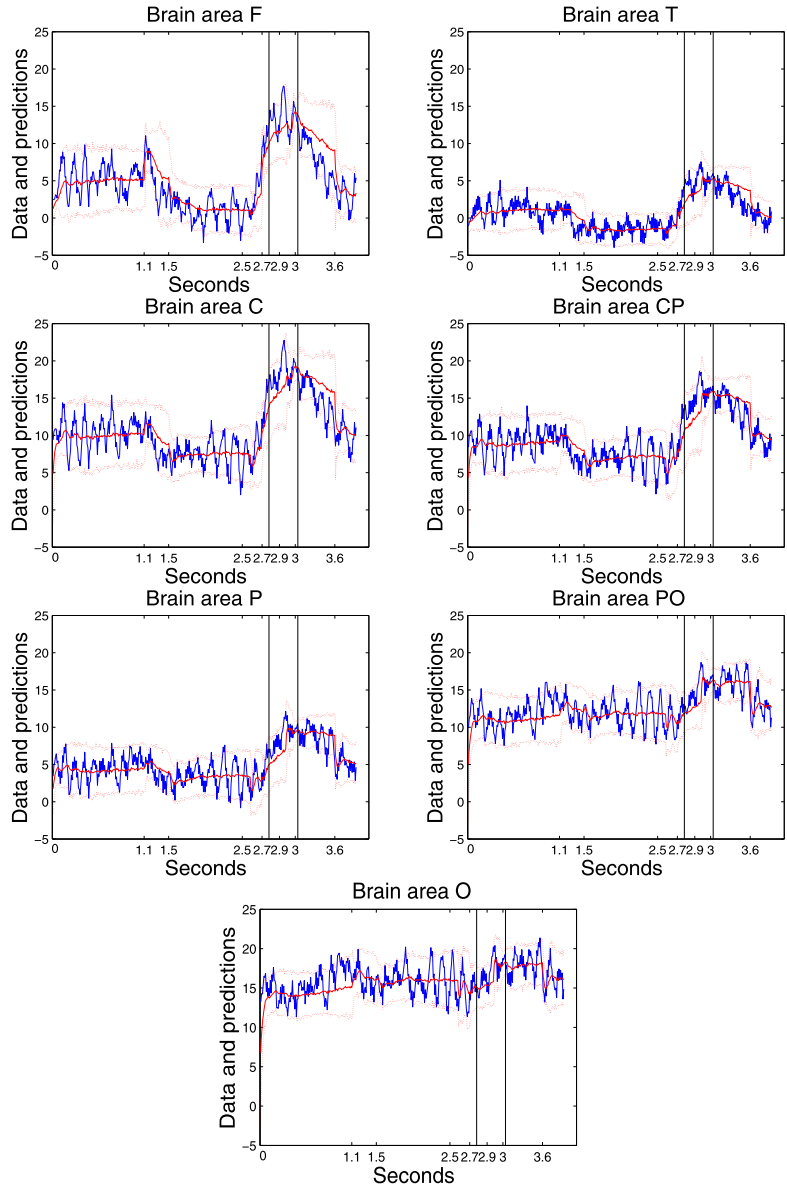}

\caption{EEG recordings (blue), one-step ahead marginal
posterior point predictions and $95\%$ posterior predictive intervals for
each brain area (red). The estimated change-point times are marked on
the horizontal axis of each plot. The two vertical lines represent
respectively the average
stimulus and response times. The brain activity is reduced roughly at
half of the initial phase of the experiment and it
increases when the cue is presented. The sharpest increases are
detected in the frontal~(F) and central (C) lobes, followed by the
central-parietal (CP), parietal (P) and temporal (T) lobes. The
estimated change in activity in the parietal-occipital (PO) and
occipital (O) areas is far less pronounced.}\label{fig2}
\end{figure}

The 32 EEG channels record neural activity arising from seven
functionally distinct brain areas, that are the frontal (F), central (C),
central-parietal (CP), parietal (P), temporal (T), parietal-occipital
(PO) and occipital (O) lobes. At each time point
the recording channels targeting each of the seven brain areas were
averaged within each trial and then across trials so as
to obtain a seven-dimensional time series.
The rationale for this preprocessing is
that recordings within each brain area exhibit similar patterns within
and across trials so that, for the purpose of our analysis, averaging
yields a lower dimensional signal less affected by
channel-specific recording noise.
These trial-averaged EEG recordings are represented in Figure~\ref{fig2}. The
activity of the different areas prior to the presentation of the
visual cue are tightly synchronized, exhibiting
oscillations of high amplitude around frequency 10~Hz and fast
low-amplitude oscillations at 60~Hz. Due to their low amplitude, the
latter are hard to see in Figure~\ref{fig2}. The lower
frequency oscillations are consistent with the so-called $\alpha$ band
reflecting eye movements. The higher frequency and lower amplitude
oscillations are due to the alternating current being used in this
experiment, suggesting an imperfect electrode grounding.

%

The trial-averaged signal at time $i$, $Y_{i}$, is modeled as $N_{7}(\mu
_{i},\Sigma_{i})$.
To derive Bayesian inferences for the mean vector and for the
covariance matrix,
we use the conjugate Normal-inverse Wishart prior:
%
\begin{eqnarray}\label{cprs}
\mu_{i} &\sim& N_{7}(\hat{\mu}_{i^{*}-1},\Sigma_{i}),\\\label{cprs2}
\Sigma_{i} &\sim& \operatorname{IW}_{7}(9,I_{7}\hat{\Sigma}_{i^{*}-1}),
\end{eqnarray}
where $1 \leq i^{*} < i$ is the time
of the last detected change-point prior to time $i$.
The marginal prior expectations are matched to the corresponding
estimated marginal posterior moments at time $i^{*}-1$ consistently with
(\ref{pmm}). At time $i$ these prior distributions are updated using
Bayes' theorem, taking into account
all data points within the interval $[i^{*}+1,i]$. Therefore, by
combining the KL
test with a static Bayesian update, this dynamic model retains a
memory of
past mean and covariance estimates from the last detected
change-point onward. From this perspective, this model can be
thought of as a form of time-varing vector autoregression which at
time $i$ has order $i-i^{*}$.

For this analysis, the initial conditions $\mu_{0}$ and $\Sigma_{0}$
were set respectively equal
to the null vector and to the identity matrix. The hyper-parameter of
the posterior density was set at
$\alpha= 0.01$, so as to detect only the most prominent changes. The
number of degrees of
freedom of the Inverse Wishart density is set so that predictive
intervals of length consistent with the set value of $\alpha$ are not
excessively inflated when a change is
detected. The distribution of the KL statistic and its value were approximated
at each time $i$ using the last 500 Gibbs sampler draws of the mean and
of the covariance matrix.

Along with the data, Figure~\ref{fig2} shows the one-step ahead marginal
posterior point predictions and their $95\%$ highest posterior
predictive intervals for
each of the seven brain areas. The estimated change-point times are
marked along the horizontal axes. The predictions
emphasize a downward shift in brain activity taking place roughly at
half of the initial phase of the experiments, followed by a sharp
increase corresponding to the cue presentation, a
downward trend following the motor response and a
stabilization of the EEG signals toward the end of the experiments.
The first two change-points identify
a transition during the first part of the experiment toward a state
of more intense attention. The third to sixth change-points
capture an abrupt increase in neural activity related to the
presentation of the visual cue, whereas the last change-point
indicates a return to a baseline activity. The sharp
increases in the activity of the frontal and central areas during the
generation of the response are consistent with their characterization
as executive and motor centers of the brain. The
intermediate increase in activity of the temporal and parietal lobes,
mainly involved in speech, hearing, memory and in the integration of
sensory inputs, reflects the mild involvement of their functions in
the execution of the task entailed by this experimental protocol. The
mild response to the visual stimulus of the parietal-occipital and
occipital areas, including the visual cortex, is somewhat surprising.
An analogous analysis of the trial-averaged EEG data from the eight
distinct channels
recording from these two areas reveals a consistently higher activity
of the occipital
channels with respect to the parietal-occipital ones but no
significant change in response to the visual stimulus.

%
\begin{figure}

\includegraphics{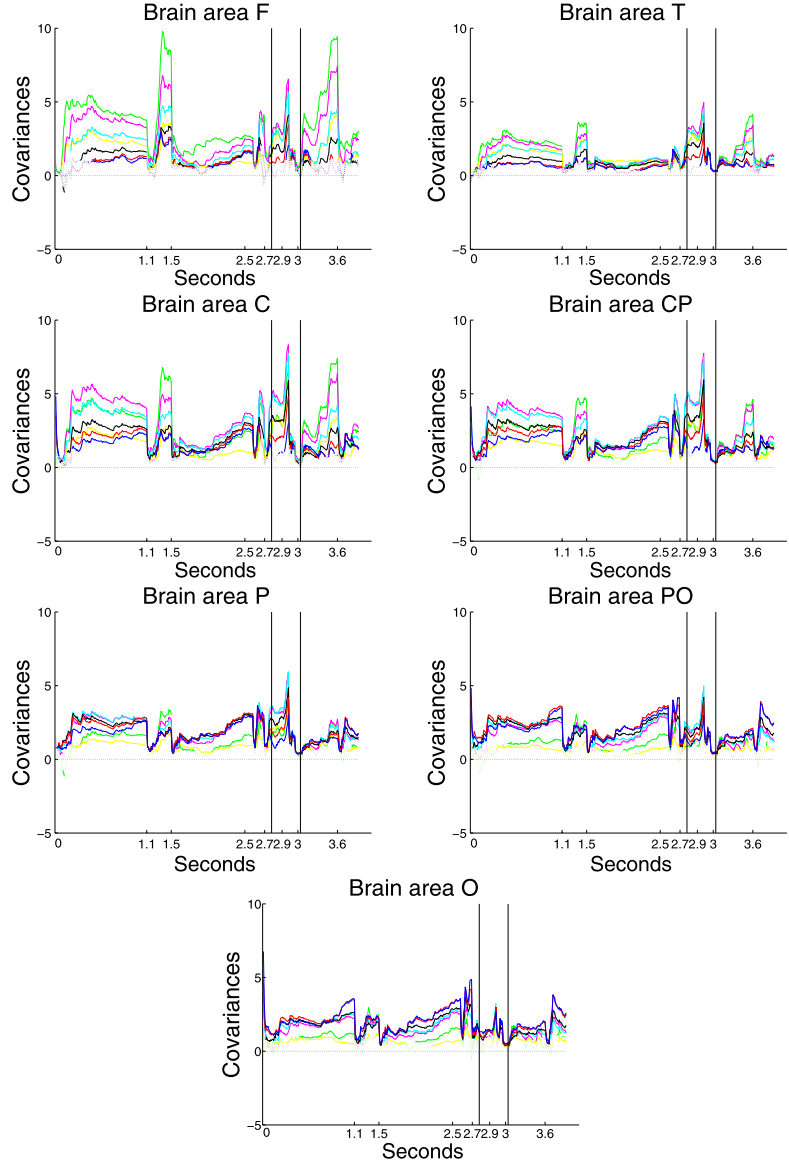}

\caption{Estimates of the time-dependent variance and
covariance functions for the frontal (green),
temporal (yellow), central (magenta), central-parietal (cyan),
parietal (black), parietal-occipital (red) and
occipital (blue) lobes. Whole segments represent
periods during which their $95\%$ posterior intervals do not
intersect zero. The estimated change-point times are marked on
the horizontal axis of each plot. The estimated covariances are almost
always positive and
time-varying, representing different levels of cooperative activity of the
seven brain areas over time. The covariance functions
are also spatially ordered, the strongest relationships being estimated
between physically adjacent brain areas.}\label{fig3}
\end{figure}

Figure~\ref{fig3} depicts the estimates of the time-dependent
variance and covariance functions for each brain area. Whole segments
represent periods during which their respective $95\%$ highest posterior
intervals do not intersect zero. The estimated change-point times are
marked on
the horizontal axis of each plot, as in Figure~\ref{fig2}. All estimated
variances and
covariances vary over time, indicating that a
time-dependent covariance matrix is an appropriate modeling
assumption for this data. The estimated covariances are almost always
positive, suggesting that the
activity of the seven brain areas is dynamically cooperative as found by
Delorme et al. (\citeyear{Delorme}). An
unexpected feature of the estimated covariance functions is their
spatial ordering over time, the strongest relationships being estimated
between adjacent
brain areas. Since neither in the Gaussian likelihood nor the priors
(\ref{cprs})--(\ref{cprs2})
include a spatial component, these estimates suggest a close
correspondence between the detected functional relationships and the
anatomical structure of the brain.

\section{Estimation of a learning curve}\label{sec3}

\begin{figure}

\includegraphics{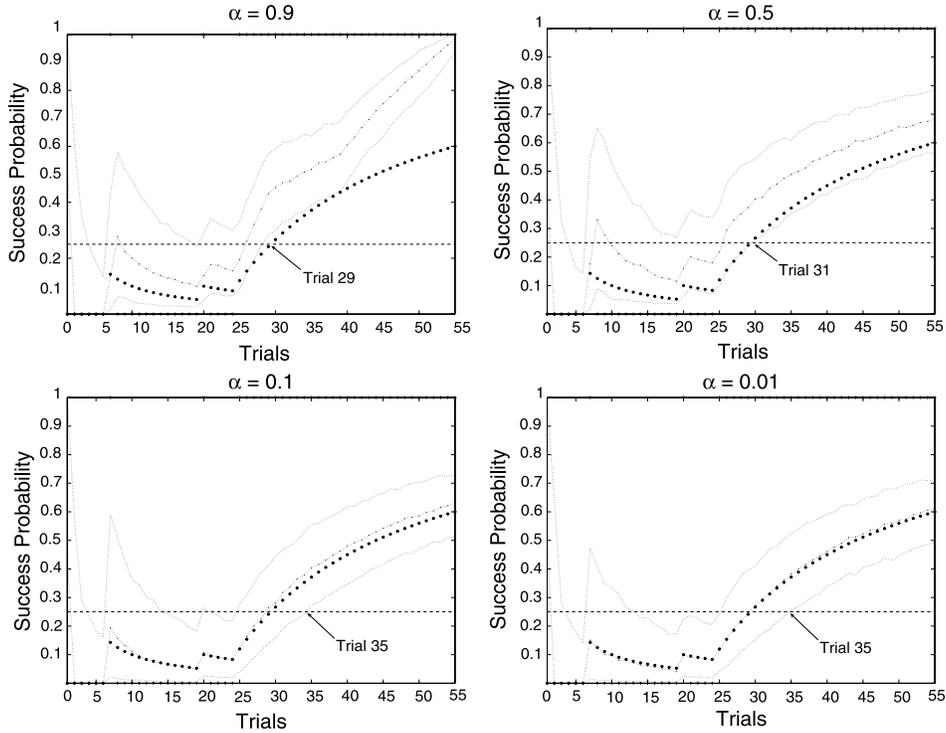}

\caption{Macaque monkey binary data and semi-parametric estimates of
their time-dependent success
probabilities using $\alpha= 0.9, 0.5, 0.1, 0.01$. The binary data
are represented as vertical ticks along the lower and upper
horizontal axes. Asterisks
represent the cumulative proportion of successful trials. The sequence
of estimates
of the success probabilities describe the macaque's learning curve
over time. The first trial at which the learning curve lies above
its chance level $0.25$, indicating that learning has effectively
taken place, is number 29. Lower values of $\alpha$ require more
extreme values of the KL statistic for detecting change-points,
making our estimates of the learning curve progressively closer to
the empirical cumulative success rates.}\label{fig4}
\end{figure}

The data analyzed in this section arises from a sequence of 55
trials during which a macaque monkey performed a location-scene
association task [Wirth et al. (\citeyear{Wirth03})]. The learning
curve is
represented by the time-dependent estimates of the trials' success
probabilities.
Smith et al. (\citeyear{Emery}) introduced a parametric state-space model
for inferring the learning performance using longitudinal
behavioral experiments. The learning curve is thereby modeled
using univariate binary time series data along with a logit link
for each trial's success probability
and a Gaussian state evolution equation for the parameters' dynamics.
In this section we use the same Bernoulli sampling
distribution for the binary trial outcomes as in
Smith et al. (\citeyear{Emery}) and we
estimate the dynamics of its success probability over time using the
semi-parametric method illustrated in Section~\ref{sec1}. A first difference
between our model and that of
Smith et al. (\citeyear{Emery}) is that we do
not use a nonlinear link function, thus imposing
fewer constraints on the shape of the learning curve. A
second difference is that the results of
Smith et al. (\citeyear{Emery}) are based on
a smoothing algorithm using both past and future data to obtain
estimates at present times, whereas our method uses past observed
values and simulated current data to update
the distribution of the success probability.
In the following analyses the success probability of the first trial
was given a uniform prior, whereas the transfer
prior (\ref{pmm}) was implemented using a conjugate Beta prior. The
data were analyzed under different values for the hyper-parameter~$\alpha$ within the range $(0.01,0.9)$, respectively requiring
from strong to weak evidence for detecting a
change-point. The distribution of the KL statistic under the null
hypothesis of no change
was approximated using ten thousand Monte Carlo samples from the Beta posterior
distribution of each trial's success probability.
For this data, the smoothed state-space estimates of
Smith et al. (\citeyear{Emery}) indicate that with $90\%$ confidence
the success
probability significantly exceeds its chance value $0.25$ from trial
23 onward, whereas their unsmoothed estimates indicate that the
chance value is significantly exceeded from trial 27 onward. Figure~\ref{fig4}
shows our estimates of the success
probabilities under the four selected values of $\alpha= 0.9, 0.5,
0.1, 0.01$. From these estimates we conclude that learning has
effectively taken place from trial 29 onward, that is, after
observing a total of 7 successes yielding an empirical cumulative
success rate of $0.24$. Figure~\ref{fig4} also compares our dynamic estimates
with the
empirical cumulative proportion of successful trials, which is
represented by asterisks. As the value of $\alpha$ decreases,
so does the number of detected change-points. In particular, under the
uniform prior for the initial success probability when $\alpha\leq
0.1$, our
estimates of the learning curve are roughly equivalent to the
empirical proportion of cumulative successes.

%

\section{Dynamic modeling of functional neuronal networks}\label{sec4}

This example illustrates the application of the method presented in
Section~\ref{sec1}
in the context of a model for networks of spiking neurons.
During the experiments analyzed here, the neural
activity of a small section of a sheep's temporal cortex is recorded
in vivo on a millisecond time frame using a multi-electrode array
[Kendrick et al. (\citeyear{Sheep})]. The goal of these experiments
was to investigate in detail the activity of brain areas
associated with memory.
Along each of 77 disconnected experiments, a sheep is shown either a
blank screen or two
images. In the latter case, a reward is given when
one of a set of ``familiar faces'' is correctly identified.
It is important to note that, even
within small brain areas, these experimental techniques only
record the activity of a relatively small fraction of neurons. Therefore,
these data do not allow reconstructing direct physical interactions
among neurons but only functional relationships among relatively
distant recording electrodes.

Introductions to the neuronal physiology and
to neuronal modeling are presented in
Fienberg (\citeyear{Fienberg}) and
Brillinger (\citeyear{Brilli2}). Recent surveys of the state-of-the-art
in multiple
spike trains modeling can be found in
Iyengar (\citeyear{Iyengar}),
Brown, Kass and Mitra (\citeyear{Brown}),
Kass, Ventura and Brown (\citeyear{KVB}),
Okatan, Wilson and Brown (\citeyear{Okatan}),
Rao (\citeyear{Rao}) and
Rigat, de Gunst and ven Pelt (\citeyear{Rigat}). Dynamic
point process neuronal models based on fully parametric state-space
representations
have been proposed by
Eden et al. (\citeyear{Eden}),
Truccolo et al. (\citeyear{Truccolo}),
Brown and Barbieri (\citeyear{BrownBarbieri}),
Srinivansan et al. (\citeyear{Srinivansan}) and
Eden and Brown (\citeyear{BrownEden}).

\subsection{Binary network model}

In what follows each element of the experimental
time series $\{Y_{i}\}_{i=1}^{77}$ is $Y_{i,k,t_{i,j(i)}}=1$ if
neuron~$k$ fires at time $t_{i,j(i)}$ during trial $i$ and
$Y_{i,k,t_{i,j(i)}} = 0$ otherwise with $j(i) = 1,\ldots,n_{i}$. We
model the joint sampling
distribution of the multiple spike train data for trial $i$, $Y_{i}$,
as a
Bernoulli process with renewal [Rigat, de Gunst and ven Pelt (\citeyear
{Rigat})].
The joint probability of a given realization $y_{i}$ is
%
\begin{eqnarray}\label{binet}
P(Y_{i} = y_{i} \vert\pi_{i}) &=& \prod_{t=t_{i,1}}^{t_{i,n_{i}}}\prod
_{k=1}^{K}\pi_{i,k,t}^{y_{i,k,t}}(1-\pi_{i,k,t})^{1-y_{i,k,t}}.
\end{eqnarray}
For model (\ref{binet}) to be biologically interpretable, the firing
probability of neuron~$k$ at time $t_{i,j(i)}$ during trial $i$,
$\pi_{i,k,t_{i,j(i)}}$, is defined as a one-to-one nondecreasing
mapping of a real-valued voltage function $v_{i,k,t_{i,j(i)}}$ onto\vspace*{1pt}
the interval $(0,1)$. The function $v_{i,k,t_{i,j(i)}}$ represents the
unnormalized difference of electrical potential across the membrane of
neuron $k$ at time $t_{i,j(i)}$.
Let $\tau_{i,k,t_{i,j(i)}}$ be the last spiking time of
neuron $k$ prior to time $t_{i,j(i)}$ during trial $i$, that is,
\begin{eqnarray*}
\tau_{i,k,t_{i,j(i)}}
= \cases{1,\qquad \mbox{if }\displaystyle\sum_{\tau=1}^{t_{i,j(i)}}Y_{i,k,\tau} = 0\mbox{ or }t_{i,j(i)}=1,\vspace*{2pt}\cr
\max\bigl\{1\leq\tau< t_{i,j(i)}{}\dvtx{} Y_{i,k,\tau} = 1\bigr\},\cr
\hspace*{32pt}\mbox{otherwise}
}
\end{eqnarray*}
and the voltage function is modeled as
%
\begin{eqnarray}\label{voltage}
v_{i,k,t_{i,j(i)}} = \sum_{l=1}^{K}\beta_{i,k,l}\sum_{w = \tau
_{i,k,t_{i,j(i)}}}^{t_{i,j(i)}-1}y_{i,l,w}.
\end{eqnarray}
The spiking probabilities are linked to (\ref{voltage}) via the
logistic mapping
\[
\pi_{i,k,t_{i,j(i)}} = \frac{e^{v_{i,k,t_{i,j(i)}}}}{1+e^{v_{i,k,t_{i,j(i)}}}}.
\]
The coefficients $\beta_{i,k,l}$ represent the strength of the
functional relationship
from neuron $l$ to neuron $k$ during trial $i$. When $\beta_{i,k,l}$
is positive during trial $i$, the firing activity of neuron $l$
promotes that of neuron
$k$, whereas when it is negative, firing of $l$ inhibits that of
$k$. When neurons $l$ and $k$ are physically connected to
each other, the coefficients $\beta_{i,k,l}$ and $\beta_{i,k,l}$
represent direct functional connections. When the two neurons
are not directly connected to each other, these network coefficients summarize
a functional relationship possibly arising from a long chain of
neurons in which activity cannot be currently recorded by the MEA technique.
The coefficients $\beta_{i,k,k}$ represent the
spontaneous spiking rate of neuron $k$ during trial $i$. The last
summation term in equation (\ref{voltage})
indicates that the membrane potential of a neuron is assumed to
be influenced only by the spiking activity of the other neurons during
its last inter-spike interval.
In this simple model we do not take into account the
occurrence of leakage currents across the neuronal membrane
[Plesser and Gerstner (\citeyear{Plesser})], so that the effect of the
spikes produced by neuron
$l$ on the voltage function does not decrease over time.

For each trial $i=1,\ldots,N$ we use a Metropolis sampler to
produce approximate posterior inferences for the $K^{2}$ model
parameters. For
each experiment, we run a neuron-wise random scan update with
independent Gaussian random walk proposals for twenty-five thousand
iterations. The initial prior for the parameters of all
experiments is Gaussian with zero mean, standard deviation
1 and zero covariance for all pairs of neurons.
Conditionally on the data $y^{0:i}$ and on the current posterior
estimates, upon observing the outcome of the $i$th${}+1$ experiment,
$y_{i+1}$, we use the
KL statistic (\ref{KL}) to test whether a significant change occurred
in any of the model's parameters. The occurrence
of such changes and the corresponding parameter estimates indicate
statistically significant variations of different aspects of the neural
activity.

\subsection{Analysis of sheep multiple spike trains}

\begin{figure}

\includegraphics{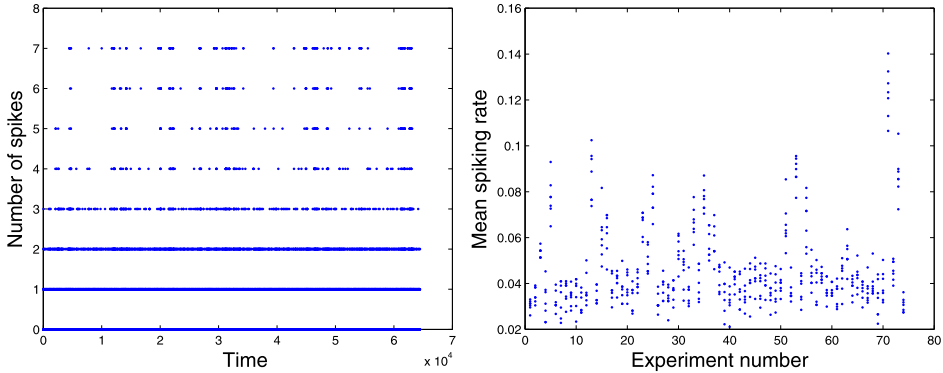}

\caption{Each dot in the left panel marks the number of recorded
spikes of the 7 most active electrodes for each millisecond
of the 77 experiments. Each dot in the right
panel marks the proportion of milliseconds during which each electrode
recorded a
spike during each experiment. The range of these mean firing rates is
0.02--0.14, reflecting the low overall spiking rates
typical for this type of recording. Clusters of points associated to
relatively high mean spiking rates suggest that the
underlying neurons may be mostly connected via
mutually excitatory functional relationships.}\label{fig5}
\end{figure}

In this section we analyze the spiking activity of the 7 most
active electrodes among the 64 recording channels.
The plot on the left in Figure~\ref{fig5} shows the number of spikes recorded
from these 7~electrodes
along all 77 experiments. The
panel on the right shows the mean spiking rates for each electrode and
experiment, which reflect the overall low spiking rates typical of
this type of measurement. The co-occurrence of relatively high
firing rates for all electrodes suggests that the most prominent
connections among
the underlying neurons may be mutually excitatory functional relationships.

%

%
\begin{table}[b]
\caption{Relative number of experiments during which both
end points of the $95\%$ posterior interval for any of the pair-wise functional
connection coefficients lie respectively above or below zero,
identifying significant
excitatory (left proportion) or inhibitory (right proportion)
relations}\label{tab1}
\begin{tabular*}{\textwidth}{@{\extracolsep{\fill}}lccccccc@{}} \hline
$\bolds{i\setminus j}$ & $\bolds{1}$ & $\bolds{2}$ & $\bolds{3}$ & $\bolds{4}$ & $\bolds{5}$ & $\bolds{6}$ & $\bolds{7}$ \\\hline
1 & $0.00$ $1.00$ & $0.10$ $0.28$ & $0.28$ $0.35$ & $0.30$ $0.12$ &$0.30$ $0.30$ & $0.08$ $0.29$ & $0.08$ $0.38$ \\
2 & $0.01$ $0.46$ & $0.00$ $1.00$ & $0.32$ $0.35$ & $0.05$ $\mathbf{0.60}$ & $0.25$ $0.25$ & $0.30$ $0.14$ & $0.41$ $0.12$ \\
3 & $0.25$ $0.13$ & $0.28$ $0.14$ & $0.00$ $1.00$ & $0.08$ $0.30$ &$0.25$ $0.12$ & $0.01$ $\mathbf{0.68}$ & $0.08$ $0.28$\\
4 & $0.09$ $0.40$ & $0.05$ $0.36$ & $0.34$ $0.14$ & $0.00$ $1.00$ &$0.33$ $0.12$ & $0.08$ $0.13$ & $0.10$ $0.36$\\
5 & $0.28$ $0.10$ & $0.08$ $0.42$ & $\mathbf{0.51}$ $0.13$ & $0.08$ $\mathbf{0.62}$ & $0.00$ $1.00$ & $0.08$ $0.28$ & $0.25$ $0.25$\\
6 & $0.10$ $0.39$ & $0.10$ $0.13$ & $\mathbf{0.63}$ $0.00$ & $0.08$ $\mathbf{0.51}$ & $0.05$ $0.30$ & $0.00$ $1.00$ & $0.08$ $0.40$\\
7 & $0.09$ $0.41$ & $0.10$ $0.14$ & $0.30$ $0.28$ & $0.08$ $0.14$ &$0.08$ $0.28$ & $0.05$ $0.39$ & $0.00$ $1.00$\\\hline
\end{tabular*}
\tabnotetext[]{tz}{The self-dependence coefficients on the main diagonal are
always found
significant and negative, representing the well-known property of neural
refractoriness. Bold entries represent functional connections which
are found significant over more than $50\%$ of the experiments. The
excitatory functional
connection from electrode 3 toward 6 is most prominent,
being significant over approximately $63\%$ of the experiments. The
most prominent inhibitory connections relate electrode 4 to 5,
which is significant over $62\%$ of the experiments, and
electrode 6 to 3, which is found significant over $68\%$ of the
experiments.}
\end{table}

\begin{figure}

\includegraphics{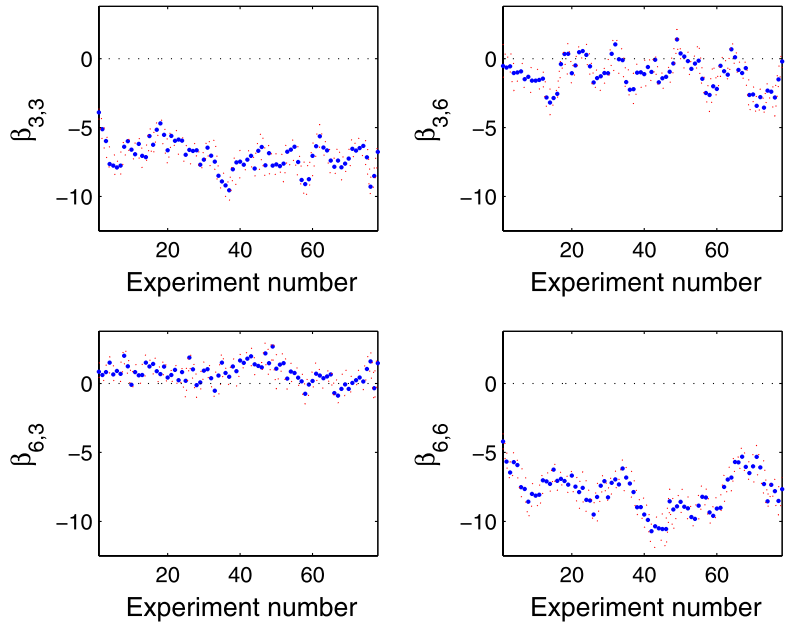}

\caption{Point estimates and $95\%$ highest
posterior intervals for the self-dependence parameters of electrodes 3
and 6 (main diagonal) and of their pair-wise functional connections
$\beta_{3,6}$ and $\beta_{6,3}$.
These two electrodes exhibit a comparable level of refractoriness over all
experiments. The estimated correlations over
experiments between the self-dependence coefficients $\beta_{3,3}$ and
those of $\beta_{3,6}$ is $-0.28$ and that between $\beta_{6,6}$ and
$\beta_{6,3}$ is $-0.29$, suggesting that neural self-inhibition may tend
to compensate for excitations and inhibitions supplied by
functionally connected cells.}\label{fig6}
\end{figure}

Table~\ref{tab1} displays summaries of the point estimates of the network
coefficients across all experiments. Each cell reports
the proportions of experiments during which each of the
network coefficients were found either significantly excitatory or
inhibitory. Significance here denotes experiments during which both
$95\%$
end points of the posterior interval of a pair-wise functional
connection lie respectively above or below zero.
The self-dependence coefficients on the main diagonal
are always found significant and negative, representing the well-known property of neural
refractoriness. The excitatory functional
connection from electrode 3 toward 6 is most prominent,
being significant over approximately $63\%$ of the experiments. The
most prominent inhibitory connections are found significant over
$62\%$ and $68\%$ of the experiments and they relate respectively
electrodes 4 to 5 and 6 to~3. Note that the time series model (\ref{voltage})
identifies a directed cyclic graph (DCG) of pair-wise functional
relationships where the connections $i \rightarrow j$ and $j
\rightarrow i$ are captured
by distinct coefficients, so that the proportion of $3 \rightarrow6$
significant excitatory connections and that of $6 \rightarrow3$
significant inhibitory connections are not constrained to add up to one.
Figure~\ref{fig6} illustrates in detail the point estimates and the $95\%$ highest
posterior intervals for the most prominent excitatory connection, $3
\rightarrow6$, together
with those of both electrodes' self-dependence and of the mostly
inhibitory connection $6 \rightarrow3$. The estimated correlation over
experiments between the self-dependence coefficients $\beta_{3,3}$ and
those of $\beta_{3,6}$ is $-0.28$ and that between $\beta_{6,6}$ and
$\beta_{6,3}$ is $-0.29$, suggesting that neural self-inhibition may tend
to compensate for excitations and inhibitions supplied by the other
recorded functionally connected cells.
%
%

%

\section{Discussion}\label{sec5}

This work is motivated by the challenges encountered in
constructing time series models when the factors driving the
dynamics of their parameters are not well understood. The
semi-parametric method illustrated here provides
flexible time-dependent estimates without relying on
explicit modeling of these dynamics. For exploratory data analyses,
such as
those presented in Sections~\ref{sec2}, \ref{sec3} and \ref{sec4}, these
estimates may suffice to address specific scientific
questions. Otherwise, appropriate measures of dependence between these
time-dependent estimates and experimental factors of interest provide a
principled basis for
more precise formulations of the parameters' dynamics. Describing the
exact form of such dependence measures is very much context-dependent
and it lies outside of the scope of this work.

A distinctive feature of the modeling approach proposed here is
that it combines elements of sequential Bayesian learning and
conditional frequentist inference along the lines of
Guttman (\citeyear{Guttman}),
Box (\citeyear{Box80}),
Berger, Brown and Wolpert (\citeyear{JimUnif}),
Meng (\citeyear{Meng}),
Gelman, Meng and Stern (\citeyear{GeMeSt96}),
Berger and Bayarri (\citeyear{JimSurprise}),
Spiegelhalter et al. (\citeyear{DIC}),
Bayarri and Morales (\citeyear{BayMor}),
Kuhnert, Mergesen and Tesar (\citeyear{Kuhnert}) and
Bayarri and Berger (\citeyear{JimInterplay}), among others.
A general treatment of such pragmatic combination of
frequentist and Bayesian ideas for model criticism can be found in Chapter
8 of
O'Hagan and Forster (\citeyear{Tony}). From this perspective, our
method is a
``Bayesianly justifiable'' procedure [Rubin (\citeyear{Rubin84})] because
only those future unobserved data that are consistent with the current
conditional posterior distribution of the model's parameters are
relevant for approximating
the distribution of the KL change-point statistic (\ref{KL}).


The latter reflects a notion of change-point as
an observation which, on the basis of the chosen model with its prior
and the observations accrued so far, is ``surprising'' from a predictive
point of view. Note that this characterization does not
depend on the parametrization of the state space nor on the
unobservable sample paths of latent states, but it
depends only on the predictives on observables.
Defining models and their properties via their one step ahead
predictive statements has been recommended, among others, by
Geisser and Eddy (\citeyear{Geisser}) and
San Martini and Spezzaferri (\citeyear{Spezza}) for predictive model
selection, by
Dawid (\citeyear{Dawid1}) in his prequential inference, by
West and Harrison (\citeyear{WH86}) for
monitoring the adequacy of Bayesian forecasting models and by
Smith (\citeyear{Jim2})
for comparing the characteristics of different forecasting models.
More recently, optimal predictive model selection criteria have
been proposed by
Barbieri and Berger (\citeyear{Maddalena}).

The results presented in Section~\ref{sec3} revealed a substantial dependence
of the estimated learning curve with respect to the
value of the hyper-parameter $\alpha$. It is important to recall
that this hyper-parameter measures how extreme a value of the KL statistic
is needed for detecting a change-point. Therefore, a dependence of its
corresponding estimated change-point process on the
value of $\alpha$ is to be expected, with lower values of this
hyper-parameter yielding less numerous change-points and vice versa.
From this perspective, our method is not meant to be fully automatic and
parameter estimates derived using different
values of $\alpha$ should be inspected to gauge their sensitivity in
the context of the
specific time series model being entertained.

In this work, a single change-point process common to all model's
parameters is used
to define their conditional posterior distribution. Should the data provide
evidence of changes of only some parameters, the
posterior distributions for the unchanging coefficients would not
make the most efficient use of the data. It is important to note that
while in principle any subset of model parameters
can be associated to a distinct change-point process, the limitations
for implementing multivariate change-point process inference within our
framework are
eminently practical. This is because marginal likelihoods for each
subset of model parameters having a different change-point process are
required to approximate the distribution of their change-point
test statistic. For classes of models where marginal likelihoods are
available in closed
form, this work can be extended by introducing a random variable
identifying groups of coefficients sharing a common change-point process.

Posterior simulation via Markov chain Monte Carlo algorithms has been
used in this work to fit multivariate time
series models and to approximate critical values of the KL
statistic. Although the current implementation of our method is
operationally realistic, these computationally intensive methods are in
fact rather impractical
for an iterative process of model formulation and criticism. Currently
two directions
are being pursued to improve the computational efficiency of our
method. On the one hand, faster resampling methods such as particle
filters [Doucet, De Freitas and Gordon (\citeyear{DoucetBook})] and
approximate Bayesian computation
[Marjoram et al. (\citeyear{TavarePNAS})] can be adopted.
Alternatively, analytical
posterior approximations can be adopted
[Tierney and Kadane (\citeyear{TierKadane})]. For
instance, in the context of sequential time series modeling,
Koyama, Perez-Bolde and Kass (\citeyear{Koyama}) recently
proposed a Laplace--Gauss posterior
approximation that obviates the use of cumbersome resampling
techniques.

\section*{Acknowledgments}
The authors acknowledge the support of the Centre for Research in
Statistical Methodology (CRiSM) at the University of Warwick and of the
Warwick Centre for Analytical Science during the development of this work.
We wish to thank Arnaud Delorme for sharing the EEG
recordings analyzed in Section~\ref{sec2}. We also wish to thank Professor
Jiangfeng Feng and collaborators for providing the
multiple spike trains analyzed in Section~\ref{sec4}.

\printaddresses

\end{document}